\documentclass[a4paper,11pt]{article}
\usepackage{pos}

\usepackage{bm}

\title{Global symmetry breaking in gauge theories: the case of multiflavor scalar
chromodynamics}
\ShortTitle{Global symmetry breaking in gauge theories}

\author*[a]{Claudio Bonati}
\emailAdd{claudio.bonati@unipi.it}

\author[b]{Andrea Pelissetto}
\emailAdd{andrea.pelissetto@roma1.infn.it}

\author[a]{Ettore Vicari}
\emailAdd{ettore.vicari@unipi.it}

\affiliation[a]{Dipartimento di Fisica dell'Universit\`a di Pisa and INFN Sezione di Pisa,\\
Largo B.~Pontecorvo 3, Pisa, Italy}

\affiliation[b]{Dipartimento di Fisica dell'Universita' di Roma ``La Sapienza'' and INFN, Sezione di Roma \\
Piazzale A.~Moro 2, Roma, Italy}

\abstract{
Universal features of continuous phase transitions can be investigated
by studying the $\phi^4$ field theory with the corresponding global 
symmetry breaking pattern. 
When gauge symmetries are present, the same technique is
usually applied to a gauge-invariant order parameter field, as in the 
Pisarski-Wilczek analysis of the QCD chiral phase transition. 
Gauge fields are thus
assumed to be irrelevant in the effective critical model, a fact that is
however far from trivial. We will investigate the validity of this approach
using three-dimensional scalar lattice models with non-abelian global and local
symmetries, for which critical exponents and scaling functions can be
numerically determined with high accuracy.
}

\FullConference{%
 The 38th International Symposium on Lattice Field Theory, LATTICE2021
  26th-30th July, 2021
  Zoom/Gather@Massachusetts Institute of Technology
}

\begin{document}
\maketitle

\section{Introduction}

In this proceeding we report on our ongoing research project aimed at a better
understanding of continuous phase transitions in three dimensional (3D) gauge
theories. It is well known that, when a spontaneous symmetry breaking is
associated with 
a continuous phase transition, universal properties emerge, which
are encoded in the $\phi^4$ field theory with the same global symmetry breaking
pattern. The main goal of our project is to understand to what extent this
approach can be applied to systems with local gauge symmetries. 

In order to explain more in detail the motivations of our work, and to
introduce the subject, it will be useful to review in some detail a specific
example likely familiar to the reader, that of the chiral phase transition in
QCD. The by now classical analysis by Pisarski and Wilczek of the finite $T$
chiral phase transition \cite{Pisarski:1983ms} goes as follows: 
\begin{enumerate}
\item first of all, we assume the transition to be continuous;
\item we model the four-dimensional finite-$T$ transition by a three-dimensional
effective field theory;
\item we assume the chiral phase transition to be described by an effective
model written using a gauge-invariant order parameter, the simplest
choice being the chiral condensate matrix;
\item we write down the most general $\phi^4$ effective Lagrangian for the
order parameter compatible with the assumed chiral symmetries, and we study its
renormalization group (RG) flow;
\item if infrared (IR) stable fixed points (FPs) of the RG-flow exist, we
conclude that the phase transition can be a continuous one, otherwise it has to
be discontinuous.
\end{enumerate}

In the first point we have to assume the phase transition to be continuous
since universality arguments can only be applied to continuous transitions, and
a discontinuous one can \emph{never} be excluded by such an argument (see also
point 5). It is instead possible to exclude (modulo the assumption 3
to be discussed in a moment) the presence of a continuous transition if no
IR-stable FPs of the effective model exist. We however have to be cautious in
drawing such a conclusion, since the absence of IR-stable FPs could also be a
consequence of the approximation scheme adopted to study the RG flow: for
example the leading order $\epsilon$-expansion computation performed in
\cite{Pisarski:1983ms} found no IR-stable FPs when the $U_A(1)$ symmetry is
explicitly broken, FPs that were instead identified by subsequent more refined
analyses (see e.~g. \cite{Pelissetto:2013hqa}).

The fundamental hypothesis in the Pisarski-Wilczek analysis is assumption
3: they assume that the effective theory is associated with a 
gauge-invariant order parameter. Of course,
 an operator has to be gauge invariant to
have a non-vanishing expectation value, e.~g., in the low-temperature 
phase (like the chiral condensate), but here the question is different: 
can the effective
theory describing the transition be defined using only gauge-invariant 
(composite) operators? If the effective Lagrangian is written by
using only gauge-invariant local composite operators, then (gauge-invariant)
gauge field correlators are noncritical, so the assumption 3 of the previous 
list is equivalent to assuming the irrelevance of the gauge modes at the
transition.  The possibility of unconventional critical points, in which gauge
fields develop critical correlations, has been put forward in the 
condensed-matter community (see, e.~g., \cite{Senthil:2003}), 
and the only way of deciding
which case is physically realized (i.e., if gauge fields are relevant or 
irrelevant) is
to compare the predictions of the universality arguments with the results of
numerical simulations or experiments.

A precise numerical estimate of the critical exponents of the chiral
phase transition in QCD is an extremely complicated task, and in some
cases even the nature of the transition (i.e., continuous or discontinuous) is
still debated, see, e.~g.,~\cite{Sharma:2019wiv}. To gain some insight into this
class of problems, it is thus convenient to switch to a simpler 
class of systems. We have considered 3D multiflavor
scalar models. For these models universality arguments analogous to those by
Pisarski and Wilczek can be carried out, and simulations can be easily
performed by using local Monte Carlo algorithms, providing precise numerical
estimates of the critical properties.

\section{The lattice model}

The basic variables of the model we will study are $N_c\times N_f$ complex
matrices $Z_{\bm x}^{cf}$, where the first index is a ``color'' index and the
second is a ``flavor'' one. For concreteness, in this proceeding we will
present results only for the simplest maximally symmetric case, whose action is
written in the form ($\mu=1,2,3$)
\begin{equation}\label{eq:Sg}
S_g =- \beta N_f \sum_{{\bm x},\mu} 
{\rm Re}\, {\rm Tr} \left[ Z_{\bm x}^\dagger \, U_{{\bm x},\hat{\mu}}
\, Z_{{\bm x}+\hat{\mu}}\right]  -
\frac{\beta_g}{N_c} \sum_{{\bm x},\mu>\nu} {\rm Re} \, {\rm Tr}\square_{{\bm x},\mu\nu}\ ,
  \quad  {\rm  Tr}\,Z_{\bm x}^\dagger Z_{\bm x} = 1\, , 
\end{equation}
where ${\bm x}$ stands for the site of a 3D cubic lattices, $U_{{\bm x},
\hat{\mu}}\in\mathrm{SU}(N_c)$ is the lattice gauge fields and the symbol
$\square_{{\bm x},\mu\nu}$ denotes the plaquette in position ${\bm x}$ laying
in the $(\mu,\nu)$ plane. The symmetry of this model is maximal, meaning that in the
ungauged limit $U_{{\bm x},\hat{\mu}}\to 1$ (i.e., for $\beta_g\to\infty$ in the
thermodynamic limit) the action is O($2N_cN_f$) symmetric, as can be seen by
writing explicitly the real and imaginary parts of $Z_{\bm x}^{cf}$. 
Note, however, that, to obtain an SU($N_c$) gauge theory,
it is sufficient to start from a scalar model with
$\mathrm{U}(N_c)\times\mathrm{U}(N_f)$ symmetry, which can be obtained by
adding to $S_g$ a quartic term proportional to ${\rm  Tr}\,(Z_{\bm
x}^\dagger Z_{\bm x})^2$; we will comment in the final section on the results
obtained when such a term is also present.

The action $S_g$ is invariant under the local transformation $Z_{\bm x}\to
G_{\bm x} Z_{\bm x}$, $U_{\bm x,\hat{\mu}}\to G_{\bm x} U_{\bm x,\hat{\mu}}
G_{\bm{x}+\hat{\mu}}^{\dag}$, where $G_{\bm x}\in$ SU($N_c$), and under the
global transformation $Z_{\bm x}\to Z_{\bm x} M$, $U_{\bm x,\hat{\mu}}\to
U_{\bm x,\hat{\mu}}$ where $M\in$ U($N_f$). The two-color case is somehow
peculiar for what concerns the global symmetry: since SU(2) is pseudo-real, it
can be shown that for $N_c=2$ the global symmetry group of $S_g$ is not
U($N_f$) but the symplectic group Sp($N_f$) (subgroup of U($2N_f$)), see
\cite{Bonati:2019zrt} for more details. 

The global symmetry U($N_f$) (or Sp($N_f$) ) of this model can be spontaneously
broken, and to identify an effective model for the transition we follow the
Pisarski-Wilczek analysis of the finite-$T$ chiral transition.  The
simplest gauge-invariant order parameter for the U($N_f$) symmetry is 
\begin{equation}\label{eq:Q}
Q_{{\bm x}}^{fg} = \sum_a \bar{Z}_{\bm x}^{af} Z_{\bm x}^{ag} - \frac{1}{N_f} \delta^{fg}\ , 
\end{equation}
which is an hermitian traceless matrix, and under the global symmetry
transforms according to $Q_{\bm x}\to M^{\dag}Q_{\bm x}M$. In fact $Q_{\bm x}$
is an order parameter for the breaking of SU($N_f$) and not U($N_f$), since it
is blind with respect to the global U(1), see \cite{Bonati:2019zrt} for a
thorough discussion of the remaining U(1) symmetry.  For the particular case
$N_c=2$ one can introduce an order parameter for Sp($N_f$) which is very
similar to $Q_{\bm x}$, see \cite{Bonati:2019zrt} for its explicit expression
and its relation to $Q_{\bm x}$ and to the U(1) symmetry.  Note that for
$N_f=1$ the order parameter $Q_{\bm x}$ identically vanishes due to the fixed
length constraint of the scalar fields; this is consistent with know rigorous
results stating that for $N_f=1$ a single thermodynamic phase is present in the
model \cite{OSFS}.

We can now write the most general Lagrangian
containing up to  4$^{\rm th}$-order powers of $Q_{\bm x}$ (more precisely of
its coarse-grained continuum counterpart $Q(\bm x)$) and invariant under the
global symmetry:
\begin{eqnarray}\label{eq:lgw}
\mathcal{L}={\rm Tr}(\partial_\mu Q)^2 
+ r \,{\rm Tr} \,Q^2 
+  w \,{\rm Tr} \,Q^3 
+  u\, ({\rm Tr} \,Q^2)^2  + v\, {\rm Tr}\, Q^4 \ .
\end{eqnarray}
For $N_f>2$, a cubic term is present, $\mathrm{Tr}\,Q^3\neq 0$, so that
a first-order phase transition is
expected. For $N_f=2$ and $N_c>3$, $\mathcal{L}$ reduces to the effective
action of the 3D O(3) universality class. Finally for $N_f=2$ and $N_c=2$ we
obtain (using the appropriate order parameter, see \cite{Bonati:2019zrt}) the
effective action of the 3D O(5) universality class; this is consistent with the
isomorphism SO(5)=Sp(2)/$\mathbb{Z}_2$. The universality class predicted by the
gauge-invariant order parameter effective Lagrangian are thus the ones reported
in Tab.~\ref{tab:univ_pred}.

\begin{table}
\centering
\begin{tabular}{l|l|l}
{}        & $N_c=2$ & $N_c>2$ \\ \hline
$N_f=2$   &  O(5) or 1$^{\mathrm{st}}$order  &  O(3) or 1$^{\mathrm{st}}$order \\ \hline
$N_f>2$   &  1$^{\mathrm{st}}$order  &  1$^{\mathrm{st}}$order
\end{tabular}
\caption{Universality class predicted by the effective model with gauge invariant order
parameter Eq.~\eqref{eq:lgw}. } \label{tab:univ_pred}
\end{table}

\section{Numerical results}

To identify the universality class of the transition of the model
in Eq.~\eqref{eq:Sg} for some values of the parameters $N_c$ and $N_f$, we
performed finite-size scaling (FSS) analyses of observables related to the
order parameter $Q_{\bm x}$ introduced in Eq.~\eqref{eq:Q}. In particular, 
using the
notation $G({\bm x}-{\bm y}) = \langle {\rm Tr}\, (Q_{\bm x} Q_{\bm y})
\rangle$ for the two-point function, we monitored the susceptibility
$\chi=\sum_{\bm x} G({\bm x})$, the second-moment finite-volume correlation
length $\xi$ and the Binder cumulant $U$, defined by
\begin{equation}
\xi^2 = \frac{1}{4 \sin^2 (\pi/L)}
\frac{\widetilde{G}({\bm 0}) - \widetilde{G}({\bm p}_m)}{\widetilde{G}({\bm p}_m)} , \quad
U = \frac{\langle \mu_2^2\rangle}{\langle \mu_2 \rangle^2} , \quad
\mu_2 = \frac{1}{V^2} \sum_{{\bm x},{\bm y}} {\rm Tr}\,Q_{\bm x} Q_{\bm y}\,,
\end{equation}
where $\tilde{G}$ denotes the Fourier transform, $L$ is the lattice size and
${\bm p}_m$ is the minimum value of the momentum consistent with the periodic
boundary conditions.

Keeping $\beta_g$ fixed, we scanned in $\beta$ the phase diagram of the model.
Since the quantities $U$ and $R_{\xi}=\xi/L$ are RG invariants, close to a
continuous transition they scale as $\approx f_{U/R_{\xi}}(X)$, where  $X =
(\beta-\beta_c)L^{1/\nu}$, $\beta_c$ is the critical value of the coupling
$\beta$, $\nu$ is the thermal critical exponent and $f_{U/R_{\xi}}$ is a
function universal up to a multiplicative rescaling of its argument (we
neglected scaling corrections for the sake of the simplicity). Since we aim to
test the predictions in Tab.~\ref{tab:univ_pred} against numerical results,
it is particularly convenient to plot $U$ as a function of $R_{\xi}$
instead of $\beta$ (or $X$): in this way we obtain the scaling law $U(\beta,L)
\approx F_U(R_\xi)$, where $F_U$ is an universal function independent of any
non-universal rescaling factor. It is thus easy to compare in a completely
unbiased way the data obtained by simulating the model in Eq.~\eqref{eq:Sg}
with those of the expected universality class.

\begin{figure}[t]
\centering
\includegraphics[width=0.48\textwidth, clip]{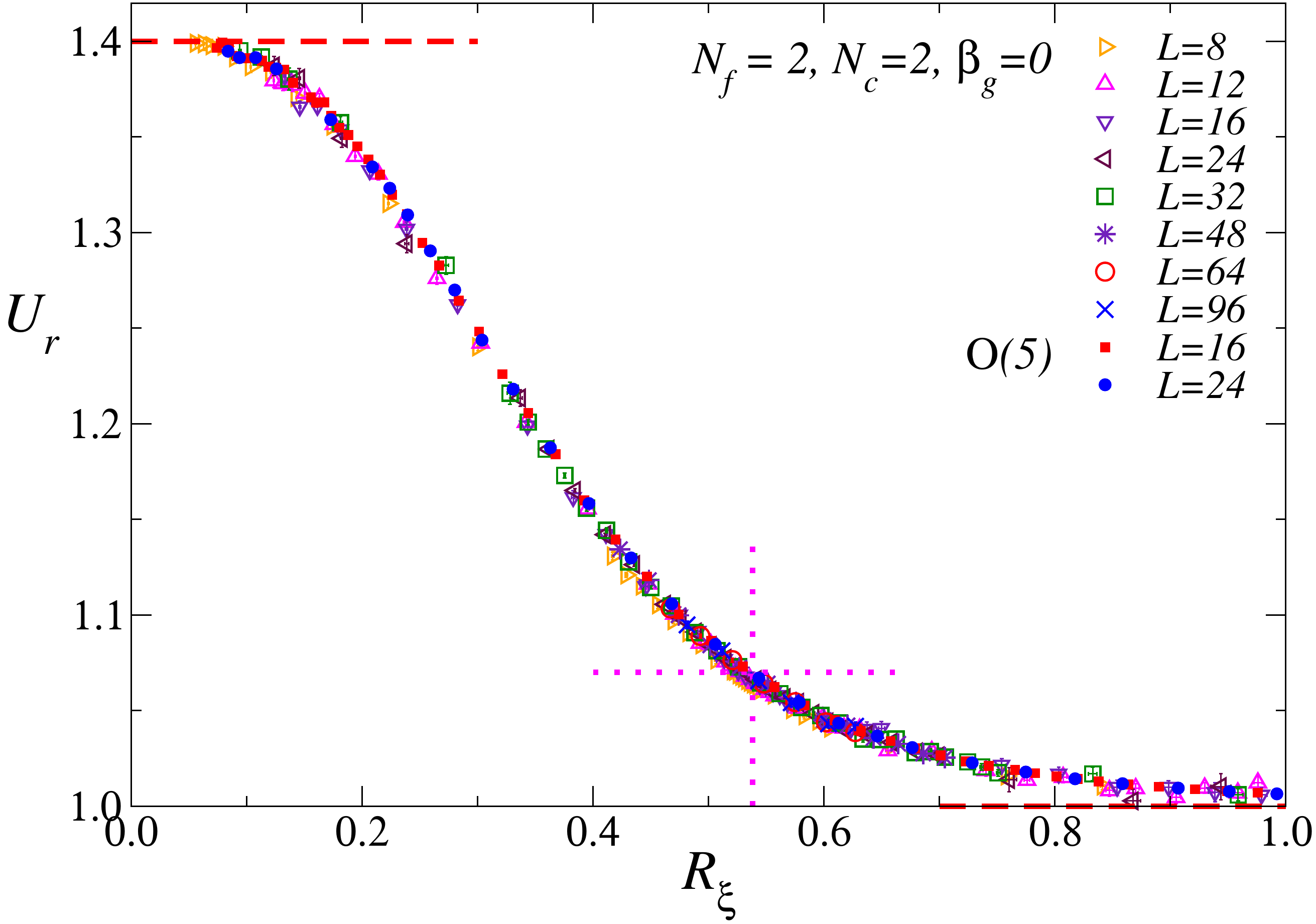}
\includegraphics[width=0.48\textwidth, clip]{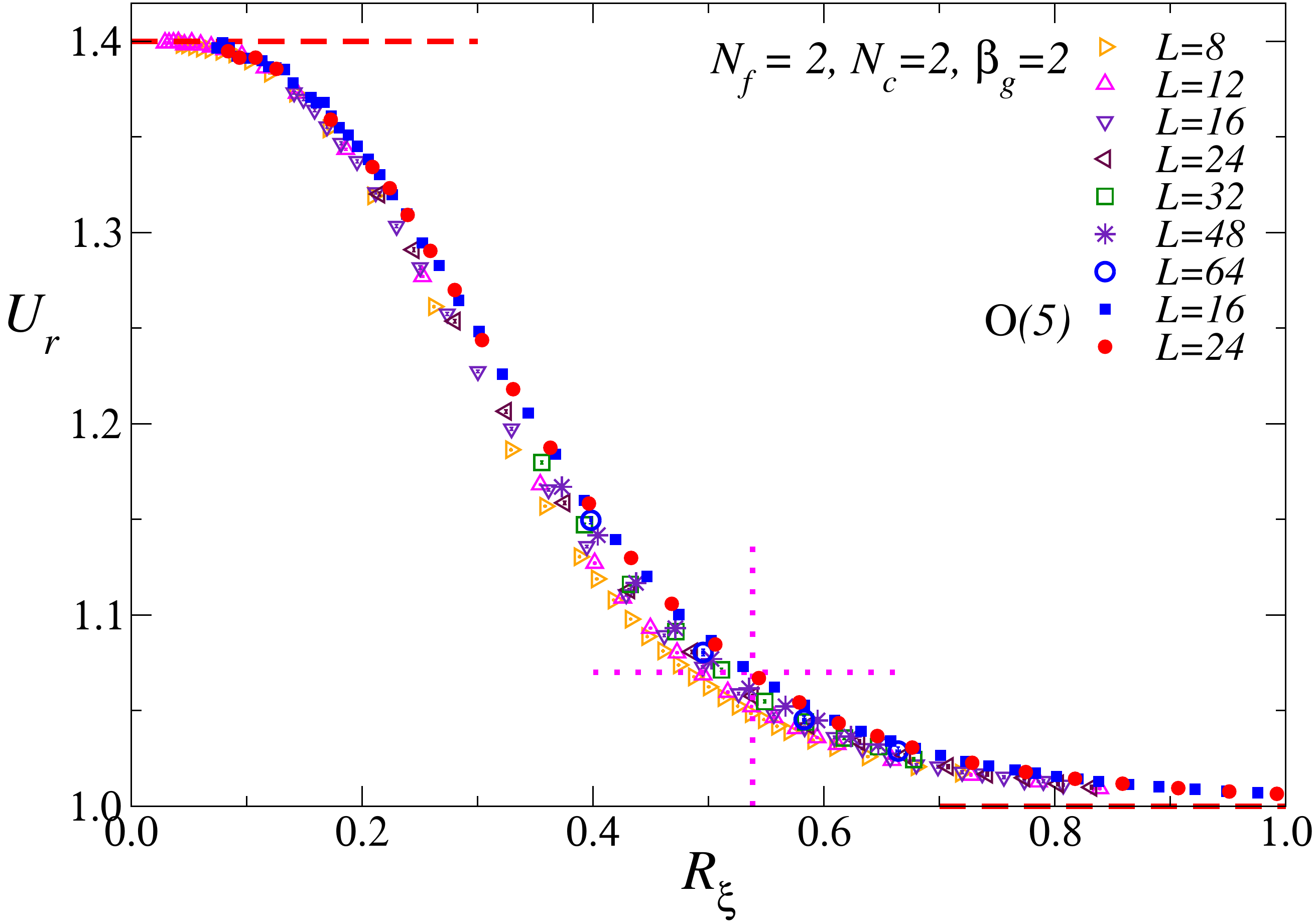}
\caption{Comparison of the scaling of the Binder cumulant (for the Sp(2) order
parameter, denoted by $U_r$) as a function of $R_{\xi}$ for the $N_c=2$,
$N_f=2$ model and the O(5) model. (left) $\beta_g=0$ and (right) $\beta_g=2$.
Dotted lines denote the known critical values of $U$ and $R_{\xi}$ for the O(5)
model.}
\label{fig:nc2scaling}
\end{figure}

\begin{figure}[b]
\centering
\includegraphics[width=0.48\textwidth, clip]{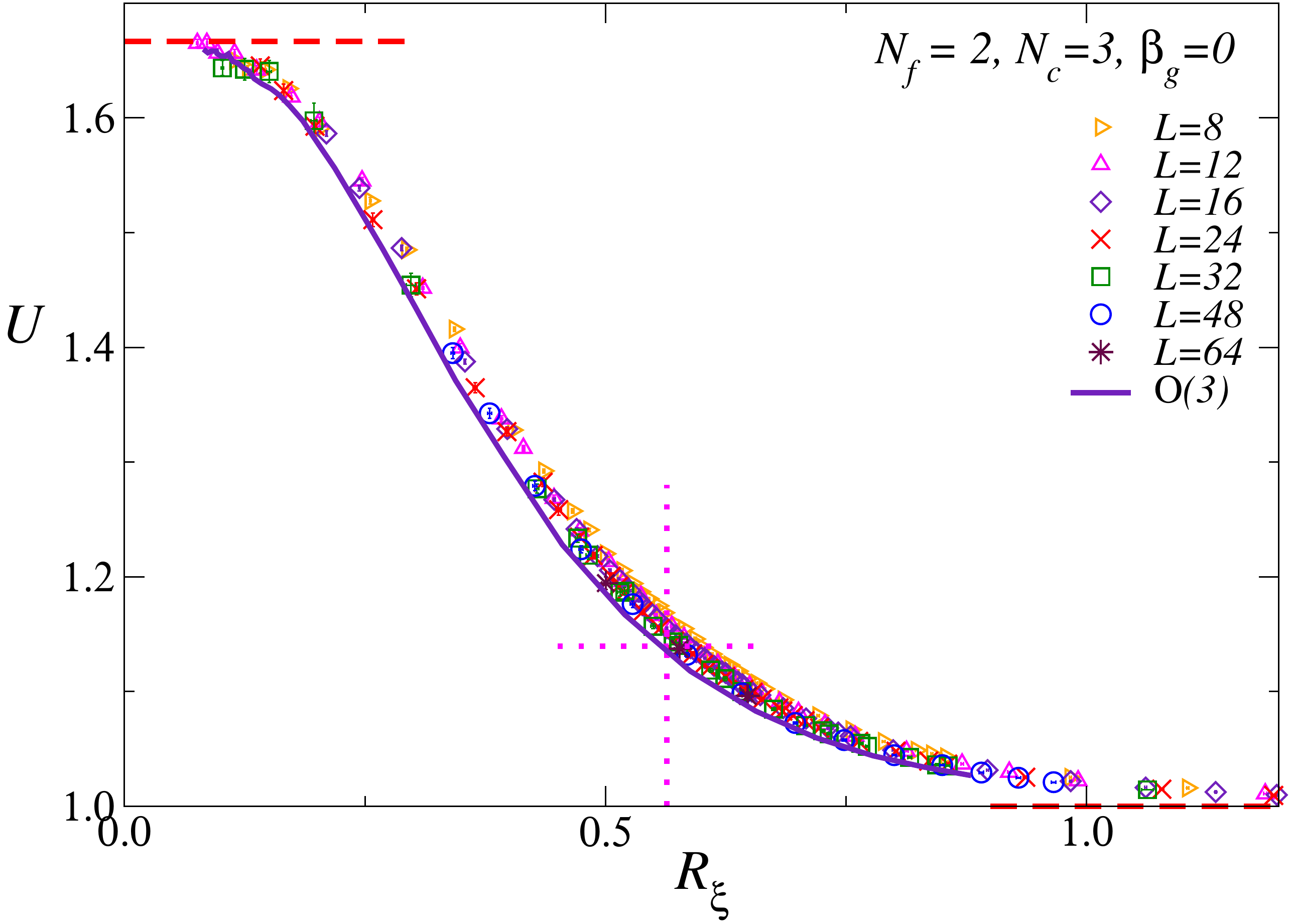}
\includegraphics[width=0.48\textwidth, clip]{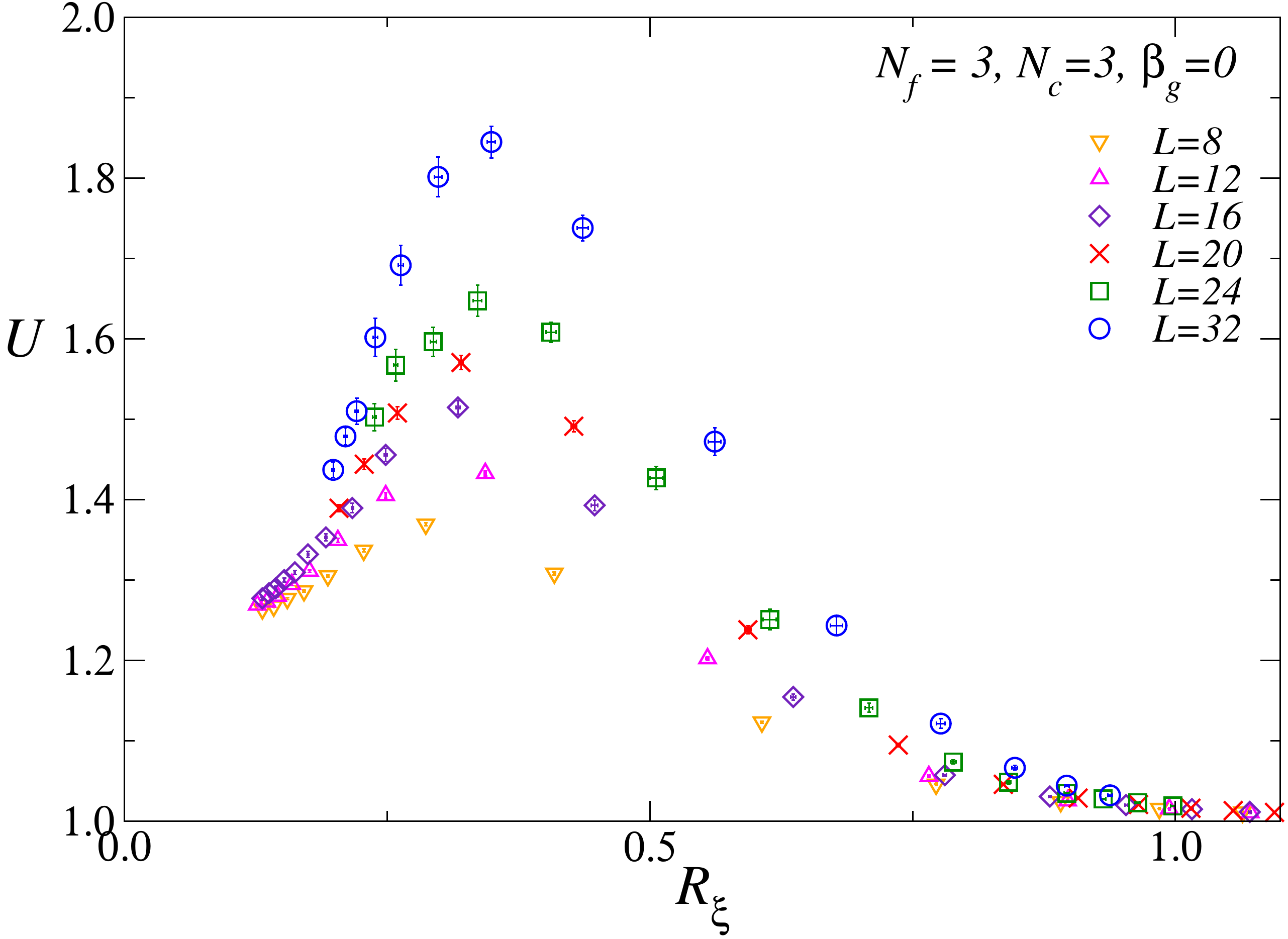}
\caption{(left) Comparison of the scaling of the Binder cumulant as a function
of $R_{\xi}$ for the $N_c=3$, $N_f=2$ model and the O(3) model.  Dotted lines
denote the known critical values of $U$ and $R_{\xi}$ for the O(3) model.
(right) Behaviour of $U$ as a function of $R_{\xi}$ for the $N_c=3$, $N_f=3$
model.}
\label{fig:nc3scaling}
\end{figure}

Such a comparison is performed in Fig.~\ref{fig:nc2scaling} between the data of
the model with $N_c=2$, $N_f=2$ (for two different values of $\beta_g$) and
those of the 3D O(5) model; note that in this case the order parameter for
Sp(2) has to be used. Both for $\beta_g=0$ and $\beta_g=2$ the data of the
gauge model approach those of the expected universality class as the lattice
size increases. An analogous comparison is reported in
Fig.~\ref{fig:nc3scaling} (left), from which we infer that the transition of
the model with $N_c=3$ and $N_f=2$ belongs to the 3D O(3) universality class, as
expected from Eq.~\eqref{eq:lgw}. More tests are discussed in
\cite{Bonati:2019zrt}, which fully support the hypothesis that the transition
of the model with $N_f=2$ is continuous and belongs to 
the O(5) and O(3) universality classes
for $N_c=2$ and $N_c=3$, respectively (for all $\beta_g$ values
investigated). Finally, for $N_f=3$ we found indications of a discontinuous
transition both for $N_c=2$ and $N_c=3$.  While the latent heat is too small to
clearly identify the typical linear dependence on the volume of the
susceptibilities, a strong indication favoring the discontinuous nature of the
transition is the absence of scaling of $U$ versus $R_{\xi}$, with $U$ that
seems to diverge as the lattice size is increased, see
Fig.~\ref{fig:nc3scaling} (right) for $N_c=3$, $N_f=3$.

\section{Conclusions and perspectives}

We discussed the universality class of the transitions of the lattice 3D
multiflavor (i.e. $N_f>1$) scalar gauge models. We first of all presented the
results of a theoretical analysis similar to the one by Pisarski and Wilczek
\cite{Pisarski:1983ms}
for the chiral phase transition in QCD, emphasizing the assumptions that are
implicit in such an analysis. We then examined the numerical lattice data
obtained by simulating the lattice action in Eq.~\eqref{eq:Sg}, which provide
compelling numerical evidence for the correctness of the predictions based on
the effective action in Eq.~\eqref{eq:lgw} for the cases studied. All the data
presented in this proceeding refers to the lattice action in
Eq.~\eqref{eq:Sg}, in which scalar fields transform according to the
fundamental representation of the gauge group and no quartic coupling is
present. 

A natural question to ask is whether the conclusions reached in this case
remain valid also in more general contexts. The type of analysis discussed in
the previous sections has been extended also to other cases, and specifically
to the case of scalar fields transforming in the adjoint representation of the
gauge group \cite{adjoint3d} and to the case in which a quartic term is also
present (both in the adjoint \cite{adjoint3d} and in the fundamental
\cite{fund3d} representation). In a subset of the region where $\beta_g$ and
the quartic coupling are positive (in the fixed-length limit negative couplings
are also allowed) something different was in fact found: in the adjoint case
with $N_c=2$ and $N_f=4$ a continuous transition incompatible with the expected
3D O(4) universality class was found \cite{adjoint3d}, while in the fundamental
case with $N_c=2$ and $N_f=40$, where Eq.~\eqref{eq:lgw} would predict a 
first-order transition, a continuous transition was identified \cite{fund3d}.

Which universality classes describe the critical behaviours associated with 
these transitions? Since they are not compatible with the predictions based on 
the effective Lagrangian written in terms of gauge-invariant operators,
Eq.~\eqref{eq:lgw}, one expects them to be associated with 
critical gauge fields.
The simplest guess for the effective action is then the \emph{continuum} field
theory for SU($N_c$) gauge fields coupled to $N_f$ scalar fields, transforming
under the appropriate representation of the gauge group. Indeed, at leading
order in the $\epsilon$-expansion, such a theory has an IR stable charged
(i.e., with a nonvanishing gauge coupling) fixed point for a large 
enough number of
scalar flavors (see \cite{adjoint3d, fund3d}).  While the identification of
these unconventional critical behaviours with the charged fixed points of the
corresponding continuum theory seems the most natural choice (also in view of
the analogous results for the U(1) gauge case \cite{chargedFP}), further
studies are required to support this identification, putting it on more solid
ground, from both the theoretical and numerical point of view. 

Finally, the relevance of these results for the case of theories with fermions,
and in particular for the finite-temperature chiral QCD transition, needs to be
better understood. A leading-order computation in the $\epsilon$-expansion
shows that the continuous field theory of SU(3) gauge fields coupled to
$N_f$ Dirac fermions has an IR-stable fixed point 
for $N_f>N_f^*=33/2$. However, these leading-order estimates 
typically largely overestimate the critical number of flavors $N_f^*$.
For instance, in the Abelian case, a charged fixed point exists for 
$N_f\gtrsim 183$, close to four dimensions, while in three dimensions 
$N_f^*$ is significantly smaller: numerically one finds
$N_f^*(3\mathrm{D})=7(2)$ (see
\cite{chargedFP}). Could it be that the continuum three-dimensional QCD has an
IR-stable fixed point relevant for the finite-temperature transition of 
four-dimensional quantum chromodynamics?


\newpage



\begin{thebibliography}{99}

\bibitem{Pisarski:1983ms}
R.~D.~Pisarski and F.~Wilczek,
``Remarks on the Chiral Phase Transition in Chromodynamics,''
Phys. Rev. D \textbf{29}, 338 (1984).

\bibitem{Pelissetto:2013hqa}
A.~Pelissetto and E.~Vicari,
``Relevance of the axial anomaly at the finite-temperature chiral transition in QCD,''
Phys. Rev. D \textbf{88}, 105018 (2013)
[arXiv:1309.5446 [hep-lat]].

\bibitem{Senthil:2003}
T.~Senthil, L.~Balents, S.~Sachdev, A.~Vishwanath and M.~P.~A.~Fisher
``Quantum criticality beyond the Landau-Ginzburg-Wilson paradigm''
Phys. Rev. B \textbf{70}, 144407 (2004)
[arXiv:cond-mat/0312617 [cond-mat.str-el]]

\bibitem{Sharma:2019wiv}
S.~Sharma,
``Recent Progress on the QCD Phase Diagram,''
PoS \textbf{LATTICE2018}, 009 (2019)
%doi:10.22323/1.334.0009
[arXiv:1901.07190 [hep-lat]].

\bibitem{Bonati:2019zrt}
C.~Bonati, A.~Pelissetto and E.~Vicari,
``Phase diagram, symmetry breaking, and critical behavior of three-dimensional lattice multiflavor scalar chromodynamics,''
Phys. Rev. Lett. \textbf{123}, 232002 (2019)
[arXiv:1910.03965 [hep-lat]];
%C.~Bonati, A.~Pelissetto and E.~Vicari,
``Three-dimensional lattice multiflavor scalar chromodynamics: interplay between global and gauge symmetries,''
Phys. Rev. D \textbf{101}, 034505 (2020)
[arXiv:2001.01132 [cond-mat.stat-mech]].

\bibitem{OSFS}
K.~Osterwalder and E.~Seiler,
``Gauge Field Theories on the Lattice,''
Annals Phys. \textbf{110}, 440 (1978);
%\bibitem{Fradkin:1978dv}
E.~H.~Fradkin and S.~H.~Shenker,
``Phase Diagrams of Lattice Gauge Theories with Higgs Fields,''
Phys. Rev. D \textbf{19}, 3682 (1979).

\bibitem{adjoint3d}
%\bibitem{Sachdev:2018nbk}
S.~Sachdev, H.~D.~Scammell, M.~S.~Scheurer and G.~Tarnopolsky,
``Gauge theory for the cuprates near optimal doping,''
Phys. Rev. B \textbf{99}, 054516 (2019)
[arXiv:1811.04930 [cond-mat.str-el]];
%\bibitem{Scammell:2019erm}
H.~D.~Scammell, K.~Patekar, M.~S.~Scheurer and S.~Sachdev,
``Phases of SU(2) gauge theory with multiple adjoint Higgs fields in 2+1 dimensions,''
Phys. Rev. B \textbf{101}, 205124 (2020)
[arXiv:1912.06108 [cond-mat.str-el]];
%\bibitem{Bonati:2021tvg}
C.~Bonati, A.~Franchi, A.~Pelissetto and E.~Vicari,
``Three-dimensional lattice SU($N_c$) gauge theories with multiflavor scalar fields in the adjoint representation,''
Phys. Rev. B \textbf{104}, 115166 (2021)
[arXiv:2106.15152 [hep-lat]].

\bibitem{fund3d}
C.~Bonati, A.~Franchi, A.~Pelissetto and E.~Vicari, 
``Phase diagram and Higgs phases of 3D lattice SU($N_c$) gauge theories with multiparameter scalar potentials,''
[arXiv:2110.01657 [cond-mat.stat-mech]].

\bibitem{chargedFP}
%\bibitem{Bonati:2020jlm}
C.~Bonati, A.~Pelissetto and E.~Vicari,
``Lattice Abelian-Higgs model with noncompact gauge fields,''
Phys. Rev. B \textbf{103}, 085104 (2021)
[arXiv:2010.06311 [cond-mat.stat-mech]];
%\bibitem{Bonati:2020ssr}
%C.~Bonati, A.~Pelissetto and E.~Vicari,
``Higher-charge three-dimensional compact lattice Abelian-Higgs models,''
Phys. Rev. E \textbf{102}, 062151 (2020)
[arXiv:2011.04503 [cond-mat.stat-mech]].


\end{thebibliography}
\end{document}